\documentclass[11pt]{article}
\usepackage{amsfonts}
\usepackage{mathrsfs}
\usepackage{graphicx}
\usepackage{amsmath}
\usepackage{amssymb}
\usepackage{subfigure}
\usepackage{epsfig}
\usepackage{graphicx}
\usepackage{color}
\usepackage{indentfirst}
\usepackage{hyperref}
\usepackage{setspace}
\usepackage{braket}
\usepackage{slashed}
\usepackage{cite}
\usepackage{multirow}
\usepackage{cancel}
\usepackage{caption}
\usepackage{array}

\hypersetup{colorlinks=true, citecolor=red, urlcolor=blue, linkcolor=blue}
\begin{document}
\title{$ \Lambda_c $ semileptonic decays}

\author{Sheng-Qi Zhang$ ^{1} $ and Cong-Feng Qiao$^{1}\footnote{Corresponding author; qiaocf@ucas.ac.cn.}$\\
\\
\normalsize{$^{1}$School of Physical Sciences, University of Chinese Academy of Sciences}\\
\normalsize{YuQuan Road 19A, Beijing 100049, China}\\
\\
}
\date{}
\maketitle

\begin{abstract}
Motivated by the recent experimental progress in the $ \Lambda_c $ decay that contains a neutron in the final state, we analyze the semileptonic decay $ \Lambda_c \rightarrow n \ell \nu_\ell $ in the framework of QCD sum rules. The transition form factors are analytically computed using three-point correlation functions and the Cutkosky cutting rules, which can be extrapolated into the physical region by employing the $ z $-series parametrization. The branching fractions of $ \Lambda_c \rightarrow n e^+ \nu_e $ and $ \Lambda_c \rightarrow n \mu^+ \nu_{\mu} $ are estimated to be $ (0.281\pm 0.056)\%$ and $ (0.275\pm 0.055)\% $, respectively. Furthermore, we calculate as well the relevant decay asymmetry observables sensitive to new physics beyond the standard model. The numerical results of semileptonic decays $ \Lambda_c \rightarrow \Lambda \ell \nu_\ell $ are also given and confronted to the latest experimental data.

\vspace{0.3cm}
\end{abstract}

\section{Introduction}
The semileptonic decay of the lightest charmed baryon $ \Lambda_c $ plays an important role in exploring strong and weak interactions in charm sectors. It can help elucidate the role of nonperturbative effects in strong interactions and provide crucial inputs for studying heavier charmed baryons and bottom baryon decay. Additionally, the precise measurement of the Cabibbo-Kobayashi-Maskawa (CKM) matrix elements $ | V_{cs} |  $ and $ | V_{cd} | $ can also provide the significant test for the standard model and the probable evidence for new physics beyond the standard model~\cite{Richman:1995wm}.

In recent years, there have been extensive measurements of the semileptonic decay modes $ \Lambda_c\rightarrow\Lambda \ell \nu_\ell $~\cite{BESIII:2015ysy, BESIII:2016ffj, BESIII:2022ysa, BESIII:2023jxv}. The most precise results of branching fractions yet are $ \mathcal{B}\left(\Lambda_c \rightarrow \Lambda e^{+} \nu_e\right)=(3.56 \pm 0.11  \pm 0.07 ) \% $~\cite{BESIII:2022ysa} and $ \mathcal{B}\left(\Lambda_c \rightarrow \Lambda \mu^{+} \nu_\mu\right)=(3.48 \pm 0.14  \pm 0.10 ) \% $~\cite{BESIII:2023jxv}, respectively. Comparing the former result with $ \Lambda_c $ inclusive semileptonic decay mode $ \mathcal{B}\left(\Lambda_c \rightarrow X e^{+} \nu_e\right)=(3.95 \pm 0.34  \pm 0.09 ) \% $~\cite{BESIII:2018mug}, it can be inferred that there still remain some potential exclusive semileptonic decay modes measurable. Recently, the BESIII collaboration reported the evidence of the decay modes containing excited states, specifically $ \Lambda_c \rightarrow \Lambda(1520) e^{+} \nu_e $ and $ \Lambda_c \rightarrow \Lambda(1405) e^{+} \nu_e $~\cite{BESIII:2022qaf}. These two decay modes yield relatively small branching fractions to be $ (1.02 \pm 0.52  \pm 0.11 ) \times 10^{-3} $ and $ (0.42 \pm 0.19  \pm 0.04 ) \times 10^{-3} $, respectively. Moreover, the measurement of two five-body semileptonic decay modes $ \Lambda_c\rightarrow \Lambda \pi^+ \pi^- e^+ \nu_e $ and $ \Lambda_c\rightarrow p K_s^0 \pi^- e^+ \nu_e $ are also performed~\cite{BESIII:2023jem}, in which the upper limits are set to be $ \mathcal{B}(\Lambda_c\rightarrow \Lambda \pi^+ \pi^- e^+ \nu_e)<3.9\times 10^{-4} $ and $ \mathcal{B}(\Lambda_c\rightarrow p K_s^0 \pi^- e^+ \nu_e)<3.3\times 10^{-4} $. In physics, besides the $ \Lambda_c $ semileptonic decay modes that include $ \Lambda(\Lambda^*) $ baryon in the final state, the exclusive semileptonic decay modes $ \Lambda_c\rightarrow n \ell \nu_\ell $ are also permitted by the standard model. However, there is still a lack of experimental data in this regard.

Theoretically, $ \Lambda_c\rightarrow n\ell\nu_\ell $ is dominated by the Cabibbo-suppressed transition $ c\rightarrow d\ell\nu_\ell $. As a result, the decay width is anticipated to be much smaller compared with the $ \Lambda_c\rightarrow \Lambda\ell\nu_\ell $ mode, which is dominated by the Cabibbo-favored transition $ c\rightarrow s\ell\nu_\ell $. Experimentally, the main challenge lies in distinguishing neutron signals from neutral noises, which leads to the problem of direct neutron detection~\cite{BESIII:2022onh, BESIII:2022xne}. Fortunately, with the improvement of detector performance and analysis technique, the BESIII collaboration has made notable progress in measuring $ \Lambda_c $ decays that involve neutron signals in the final state~\cite{BESIII:2022onh, BESIII:2022xne, BESIII:2016yrc, BESIII:2022bkj}. It is predictable that the experimental data for the decay mode $ \Lambda_c\rightarrow n\ell\nu_\ell $ will be available in the near future, making it beneficial to explore this process theoretically. Furthermore, the semileptonic decay $ \Lambda_c\rightarrow n \ell\nu_\ell $ is an exceptional candidate for extracting the magnitude of the CKM matrix element $ |V_{cd}| $. Currently, the determination of $ |V_{cd} |$ relies primarily on the charm meson semileptonic decay $ D\rightarrow \pi \ell \nu_\ell $~\cite{Belle:2006idb, CLEO:2009svp, BaBar:2014xzf, BESIII:2015tql, BESIII:2017ylw}. Therefore, it is of great importance to investigate the semileptonic decay $ \Lambda_c\rightarrow n \ell\nu_\ell $ both experimentally and theoretically since such studies are crucial for providing precise verification for $ | V_{cd} | $ in the charm baryon sector.

In the past, theoretical investigations for $ \Lambda_c\rightarrow n \ell \nu_\ell$ semileptonic decay have been performed in-depth in a variety of methods, such as the light-cone sum rules (LCSR)~\cite{Li:2016qai, Azizi:2009wn, Khodjamirian:2011jp}, the light front approach (LF)~\cite{Zhao:2018zcb}, the covariant confined quark model (CCQM)~\cite{Gutsche:2014zna}, the constituent quark model (CQM)~\cite{Pervin:2005ve}, the relativistic quark model (RQM)~\cite{Faustov:2016yza}, the $SU(3)$ flavor symmetry~\cite{Lu:2016ogy, Geng:2019bfz}, the MIT bag model (MBM)~\cite{Geng:2020fng}, and the lattice QCD (LQCD)~\cite{Meinel:2017ggx}. Additionally, QCD sum rules (QCDSR) has also been widely utilized to deal with the baryonic decay mode~\cite{Dai:1996xv, Dosch:1999pr, Huang:1998rq, Huang:1998ek, MarquesdeCarvalho:1999bqs, Shi:2019hbf, Zhao:2020mod, Zhao:2021sje, Xing:2021enr}. Rather than a phenomenological model, QCDSR is a QCD-based theoretical framework that systematically incorporates nonperturbative effects at each dimension. To evaluate the form factors in the weak transitions, the three-point correlation functions are constructed with appropriate interpolating currents. After equating two representations of the three-point correlation functions, i.e., the QCD representation and the phenomenological representation, using quark-hadron duality, the form factors will be formally determined. In this work, we will apply QCDSR to calculate the form factors of the $ \Lambda_c \rightarrow n \ell \nu_\ell $ semileptonic decay mode, after which the branching fractions as well as some other relevant decay asymmetry observables are also obtained. Besides, the numerical results of $ \Lambda_c \rightarrow \Lambda \ell \nu_\ell $ semileptonic decay are also given and compared with the latest experimental results.

The rest of the paper is structured as follows: in Sec.~\ref{Formalism} we interpret the basic idea of QCDSR for the three-point correlation functions. The numerical results and analysis are presented in Sec.~\ref{Numerical}. The conclusions and discussions are given in the last section.

\section{Formalism}\label{Formalism}
\begin{figure}[ht]
\centering
\includegraphics[width=7.5cm]{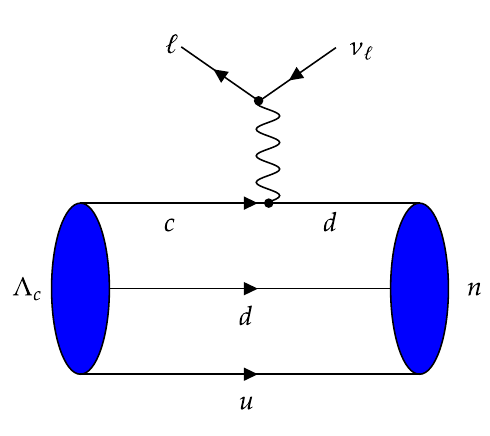}
\caption{Feynman diagram for $ \Lambda_c \rightarrow n \ell \nu_\ell$ semileptonic decay.}
\label{fig:diagram}
\end{figure}

The $ \Lambda_c \rightarrow n \ell \nu_\ell$ decay is dominated by the Cabibbo-suppressed transition $ c\rightarrow d \ell \nu_\ell $ at the quark level. The effective Hamiltonian depicting this transition is written as
\begin{align}
\mathcal{H}_{e f f}=\frac{G_F}{\sqrt{2}} V_{c d} [\bar{\ell} \gamma_\mu\left(1-\gamma_5\right) \nu_{\ell}] [\bar{d} \gamma^\mu\left(1-\gamma_5\right) c],
\end{align}
where $ G_F $ denotes the Fermi constant and $ V_{c d} $ is the CKM matrix element. The Feynman diagram of $ \Lambda_c \rightarrow n \ell \nu_\ell$ is shown in Fig.~\ref{fig:diagram}. The leptonic part of this decay mode can be obtained through electroweak perturbation theory, while the hadronic part cannot be calculated perturbatively due to its involvement in the low-energy aspects of QCD. In general, the weak transition matrix element of the hadronic part can be parametrized in terms of transition form factors
\begin{align}
\label{matrix}
\bra{\Lambda_c(q_1)}j_\mu\ket{n(q_2)}&=\bar{u}_{\Lambda_c}(q_1) \bigg[f_1(q^2) \gamma_\mu+i f_2(q^2)\sigma_{\mu\nu}\frac{q^\nu}{M_{\Lambda_c}}+f_3(q^2) \frac{q_\mu}{M_{\Lambda_c}} \bigg] u_n(q_2)\nonumber\\
&-\bar{u}_{\Lambda_c}(q_1)\bigg[g_1(q^2) \gamma_\mu+i g_2(q^2) \sigma_{\mu\nu}\frac{q^\nu}{M_{\Lambda_c}}+g_3(q^2) \frac{q_\mu}{M_{\Lambda_c}}\bigg]\gamma_5 u_n(q_2),
\end{align}
where $ q_1 $ and $ q_2 $ are the four-vector momentum of the initial state $ \Lambda_c $ and final state neutron, respectively. The momentum transfer $ q $ is defined as $ q=q_1-q_2 $.

To calculate the transition form factors by QCDSR, the three-point correlation functions can be formally constructed as
\begin{equation}
\label{3pcf}
\Pi_\mu(q_1^2, q_2^2, q^2)=i^2 \int d^4 x\; d^4 y\; e^{i(q_1 x-q_2 y)} <0|T\{j_{\Lambda_c}(x)j_\mu(0)j^{\dagger}_{n}(y) \}|0>.
\end{equation}
The weak transition current $ j_\mu $ is defined as $ j_\mu=\bar{c} \gamma_\mu\left(1-\gamma_5\right) d $, and the interpolating currents of $ \Lambda_c $ and the neutron take the following quark structure~\cite{Zhao:2020mod, Chung:1981wm}:
\begin{align}
\label{current-lc}
j_{\Lambda_{c}}&=\epsilon_{i j k}\left(u_i^T C \gamma_5 d_j\right) c_k, \\
\label{current-n}
j_n&=\epsilon_{i j k}\left(u_i^T C \gamma_5 d_j\right) d_k,
\end{align}
where the subscripts $ i $, $ j $, and $ k $ represent the color indices and $ C $ is the charge conjugation matrix. It should be mentioned that there are other choices of interpolating currents of the neutron, such as Ioffe type and tensor type~\cite{Ioffe:1982ce, Ioffe:1981kw, Braun:2006hz}. However, as stated in Ref.~\cite{Huang:1998rq}, if the approximations in sum rule calculations are justified to be good enough, these different currents should give roughly the same physical results.

On the phenomenological side, after inserting a complete set of intermediate hadronic states and exploiting double dispersion relations, the three-point correlation functions in Eq.~(\ref{3pcf}) can be described as,
\begin{align}
\Pi_\mu^{\text{phe}}(q_1^2, q_2^2, q^2)=&\sum_{\text{spins}}\frac{\bra{0}j_{\Lambda_c}\ket{\Lambda_c(q_1)}\bra{\Lambda_c(q_1)}j_\mu\ket{n(q_2)}\bra{n(q_2)}j_{n}\ket{0}}{(q_1^2-M_{\Lambda_c}^2)(q_2^2-M_{n}^2)}\nonumber\\
+&\text{higher resonances and continuum states},
\end{align}
where $ M_{\Lambda_c} $ and $ M_{n} $ denote the mass of $ \Lambda_c $ and the neutron, respectively. The vacuum-to-baryon transition amplitudes can be parametrized by defining the decay constants,
\begin{align}
\bra{0}j_{\Lambda_c}\ket{\Lambda_c(q_1)}&=\lambda_{\Lambda_c}u_{\Lambda_c}(q_1),\\
\bra{0}j_{n}\ket{n(q_2)}&=\lambda_{n}u_{n}(q_2),
\end{align}
where $ \lambda_{\Lambda_c} $ and $ \lambda_{n} $ represent the decay constants of $ \Lambda_c $ and the neutron, respectively. By introducing the hadronic transition matrix elements in Eq.~(\ref{matrix}) and utilizing the spin sum completeness relations, $ \sum u_{\Lambda_c}(q_1)\bar{u}_{\Lambda_c}(q_1)=\slashed{q}_1+M_{\Lambda_c} $ and $ \sum u_{n}(q_2)\bar{u}_{n}(q_2)=\slashed{q}_2+M_{n} $, we can finally obtain the phenomenological representation of the three-point correlation functions of Eq.~(\ref{3pcf}),
\begin{align}
\label{3ptphe}
\Pi_\mu^{\text{phe}}(q_1^2, q_2^2, q^2) & =\frac{\lambda_{n} \left(\slashed{q}_2+M_{n}\right)\bigg[f_1(q^2) \gamma_\mu+i f_2(q^2)\sigma_{\mu\nu}\frac{q^\nu}{M_{\Lambda_c}}+f_3(q^2)\frac{q_\mu}{M_{\Lambda_c}}\bigg]\lambda_{\Lambda_c}\left(\slashed{q}_1+M_{\Lambda_c}\right)}{(q_1^2-M_{\Lambda_c}^2)(q_2^2-M_{n}^2)} \nonumber\\
&-\frac{\lambda_{n} \left(\slashed{q}_2+M_{n}\right)\bigg[g_1(q^2) \gamma_\mu+i g_2(q^2)\sigma_{\mu\nu}\frac{q^\nu}{M_{\Lambda_c}}+g_3(q^2)\frac{q_\mu}{M_{\Lambda_c}} \bigg]\gamma_5\lambda_{\Lambda_c}\left(\slashed{q}_1+M_{\Lambda_c}\right)}{(q_1^2-M_{\Lambda_c}^2)(q_2^2-M_{n}^2)}.
\end{align}
It should be noted that we assume $ f_3(q^2) $ and $ g_3(q^2) $ to be negligible in this study as they will contribute to semileptonic decays at $\mathcal{O}\left(m_{\ell}^2 \right)$~\cite{Emmerich:2016jjm, MarquesdeCarvalho:1999bqs}.

\begin{figure}[ht]
\centering
\includegraphics[width=13cm]{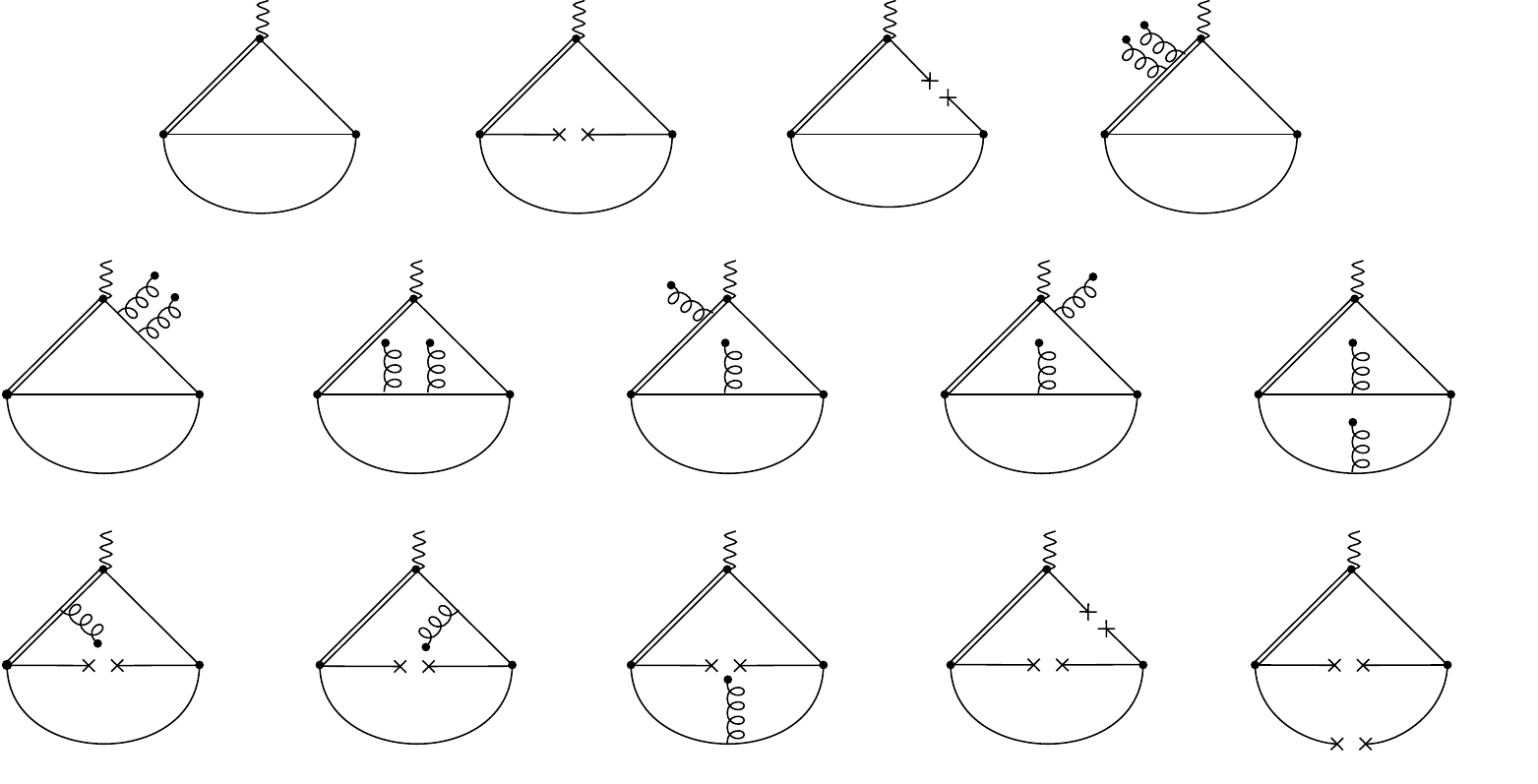}
\caption{The Feynman diagrams for the computation of $ \rho^{\text{QCD}}_\mu(s_1,s_2,q^2) $. Double solid lines represent charm quark, and ordinary solid lines denote light quark.}
\label{fig:condensate}
\end{figure}

On the QCD side, the three-point correlation functions of Eq.~(\ref{3pcf}) can be expressed by operator-product expansion (OPE) and double dispersion relations,
\begin{align}
\Pi_\mu^{\text{QCD}}(q_1^2, q_2^2, q^2)=\int_{s_1^\text{min}}^{\infty}d s_1\int_{s_2^\text{min}}^{\infty}d s_2\frac{\rho^{\text{QCD}}_\mu(s_1,s_2,q^2)}{(s_1-q_1^2)(s_2-q_2^2)},
\end{align}
where $ s_{1(2)}^\text{min} $ is the kinematic limit. $ \rho^{\text{QCD}}_\mu(s_1,s_2,q^2) $ stands for the spectral density, which can be obtained through the application of Cutkosky cutting rules~\cite{Zhao:2020mod, MarquesdeCarvalho:1999bqs, Shi:2019hbf, Zhao:2021sje, Xing:2021enr, Wang:2012hu, Yang:2005bv, Du:2003ja}. In this work, contributions up to dimension 6 are considered in $ \rho^{\text{QCD}}_\mu(s_1,s_2,q^2) $, which can be expressed as:
\begin{align}
\rho^{\text{QCD}}_\mu(s_1,s_2,q^2)&=\rho^{\text{pert}}_\mu(s_1,s_2,q^2)+\rho^{\langle \bar{q}q\rangle}_\mu(s_1,s_2,q^2)+\rho^{\langle g_s^2G^2 \rangle}_\mu(s_1,s_2,q^2)\nonumber\\
&+\rho^{\langle g_s \bar{q}\sigma \cdot  G q \rangle}_\mu(s_1,s_2,q^2)+\rho^{\langle \bar{q}q \rangle^2}_\mu(s_1,s_2,q^2).
\end{align}
The first term corresponds to the perturbative contribution, while $ \langle\bar{q}q \rangle$, $  \langle g_s^2G^2 \rangle $, $ \langle g_s \bar{q}\sigma \cdot  G q \rangle $, and $ \langle \bar{q}q \rangle^2 $ represent condensates that describe the nonperturbative effects. The relevant Feynman diagrams are plotted in Fig.~\ref{fig:condensate}.

To establish the relation between phenomenological representation and QCD representation, the quark-hadron duality is adopted,
\begin{equation}\label{quark-hadron}
\Pi_\mu^{\text{phe}}(q_1^2, q_2^2, q^2)\simeq\int_{s_1^\text{min}}^{s_1^0}d s_1\int_{s_2^\text{min}}^{s_2^0}d s_2\frac{\rho^{\text{QCD}}_\mu(s_1,s_2,q^2)}{(s_1-q_1^2)(s_2-q_2^2)}.
\end{equation}
Here, $ s_1^0 $ and $ s_2^0 $ denote the threshold parameters of $ \Lambda_c $ and the neutron, respectively. After taking into account the double Borel transform to suppress the higher excited states and continuum states contributions, the analytic expression of $ f_i(q^2) $ and $ g_i(q^2) $ can be derived:
\begin{align}
\label{f1}
f_1(t)=g_1(t)=&\frac{e^{M_{\Lambda_c}^2/M_{B_1}^2}e^{M_{n}^2/M_{B_2}^2}}{\lambda_{\Lambda_c}\lambda_{n}M_{\Lambda_c}}\bigg[\int_{s_1^\text{min}}^{s_1^0}d s_1\int_{s_2^\text{min}}^{s_2^0}d s_2\int d \xi \frac{3\,m_c\,\xi}{64\,\pi^4\lambda(s_1,s_2,t)^{3/2}}\times\nonumber\\
&\big(m_c^2(s_1-t-s_2)-t\,(s_1-t+s_2-2\,\xi) \big)\times e^{-s_1/M_{B_1}^2}e^{-s_2/M_{B_2}^2}\nonumber\\
&+\frac{m_c\langle \bar{q}q\rangle^2}{6}e^{-m_c^2/M_{B_1}^2}e^{-m_d^2/M_{B_2}^2}\bigg],\\[10pt]
\label{f2}
f_2(t)=g_2(t)=&\frac{e^{M_{\Lambda_c}^2/M_{B_1}^2}e^{M_{n}^2/M_{B_2}^2}}{\lambda_{\Lambda_c}\lambda_{n}}\int_{s_1^\text{min}}^{s_1^0}d s_1\int_{s_2^\text{min}}^{s_2^0}d s_2\int d \xi \frac{3\,\xi}{64\,\pi^4\lambda(s_1,s_2,t)^{5/2}}\Big[\nonumber\\
&m_c^4 \big(s_2(s_1+t)+(s_1-t)^2-2\,s_2^2\big)-\nonumber\\
&m_c^2 \big(s_1^3-s_1^2 (t+s_2+2 \,\xi)-s_1 (t^2+2 \,t \,(\xi-3 \,s_2)+s_2 (s_2-4 \,\xi))\nonumber\\
&+(t-s_2) (t^2+4 \,t \,\xi-s_2 (s_2-2 \,\xi)) \big)-\nonumber\\
&t \big(-2\xi (-2 \,s_1^2+s_1 (t+s_2)+(t-s_2)^2)-3 \,\xi^2 (s_1+t-s_2)\nonumber\\
&-s_1 s_2 (s_1+t)-s_1 (s_1-t)^2+2 \,s_1 s_2^2\big)\Big]\times e^{-s_1/M_{B_1}^2}e^{-s_2/M_{B_2}^2},
\end{align}
where we define $ t = q^2 $ and $ \lambda(s_1,s_2,t) = s_1^2+s_2^2+t^2-2 s_1 s_2-2 s_1 t-2 s_2 t $. $ M_{B_1}^2 $ and $ M_{B_2}^2 $ represent the Borel parameters which will appear after double Borel transform. Note, there are twenty-four Lorentz structures of form factors in Eq.~(\ref{quark-hadron}). Here, we only present those that meet the parameter selection criteria. Details will be discussed in the next section. The variable $ \xi $ is introduced in the integral through the phase space integration by utilizing Cutkosky cutting rules. It can be observed from Eqs.~(\ref{f1}) and (\ref{f2}) that the quark condensate $ \langle \bar{q}q\rangle $ and the mixed quark-gluon condensate $ \langle g_s \bar{q}\sigma \cdot  G q \rangle $ do not contribute to the transition form factors, while we find the contribution from gluon condensate $\langle g_s^2G^2 \rangle$ is negligible and can be ignored. Thus, only the four-quark condensate $ \langle \bar{q}q\rangle^2 $ determines the primary nonperturbative contribution to $ f_1 $, which is in agreement with previous theoretical studies of heavy to light transitions~\cite{MarquesdeCarvalho:1999bqs, Huang:1998rq, Huang:1998ek}.

\section{Numerical results and discussions}\label{Numerical}
In our numerical calculation, the following input parameters are adopted~\cite{Shifman:1978by, Colangelo:2000dp, Du:2003ja, Yang:2005bv, ParticleDataGroup:2022pth, Khodjamirian:2011jp, Chung:1984gr, Wan:2021vny},
\begin{align}
\label{parameter}
\langle\bar{q}q\rangle &= -(0.24 \pm 0.01)^3 \text{GeV}^3, \nonumber \\
s_1^0 &= (9.5 \sim 10.5) \,\text{GeV}^2, \quad s_2^0 = (2.4\sim 3.0) \,\text{GeV}^2\nonumber\\
m_c &= 1.27 \pm 0.02 \,\text{GeV}, \quad m_d = 4.67^{+0.48}_{-0.17} \,\text{MeV}, \nonumber \\
\lambda_{\Lambda_c} &= 0.0119 \,\text{GeV}^3, \quad \lambda_{n} = 0.02 \,\text{GeV}^3,\nonumber\\
M_{\Lambda_c} &= 2.286 \,\text{GeV}, \quad M_n = 0.938 \,\text{GeV}.
\end{align}	
Here, the standard value of the quark condensate $ \langle \bar{q}q \rangle$ is taken at the renormalization point $ \mu = 1\,\text{GeV} $. The decay constants and the threshold parameters are determined using the two-point sum rules~\cite{Khodjamirian:2011jp, Chung:1984gr}, employing the same interpolating currents of Eq.~(\ref{current-lc}) and Eq.~(\ref{current-n}).

Moreover, two additional free parameters, namely the Borel parameters $ M_{B_1}^2 $ and $ M_{B_2}^2 $, are introduced in the framework of QCDSR. For simplicity, we adopt the following relation of Borel parameters~\cite{Shi:2019hbf,MarquesdeCarvalho:1999bqs, Leljak:2019fqa},
\begin{align}
\frac{M_{B_1}^2}{M_{B_2}^2}=\frac{M_{\Lambda_c}^2-m_c^2}{M_{n}^2-m_d^2}.
\end{align}

In general, two criteria are employed to determine the values of Borel parameters. First is the pole contribution. In order to investigate the contribution of ground-state hadrons, the pole contribution has to dominate the spectrum. Thus, the pole contribution can be selected larger than $ 40\% $ for the transition form factors, which can be formulated as follows:
\begin{align}
R^{\text{PC}}_{\Lambda_c}=\frac{\int_{s_1^\text{min}}^{s_1^0}d s_1\int_{s_2^\text{min}}^{s_2^0}d s_2}{\int_{s_1^\text{min}}^{\infty}d s_1\int_{s_2^\text{min}}^{s_2^0}d s_2},\\[5pt]
R^{\text{PC}}_{n}=\frac{\int_{s_1^\text{min}}^{s_1^0}d s_1\int_{s_2^\text{min}}^{s_2^0}d s_2}{\int_{s_1^\text{min}}^{s_1^0}d s_1\int_{s_2^\text{min}}^{\infty}d s_2}.
\end{align}
These two ratios can be regarded as the pole contribution from the $ \Lambda_c $ channel and neutron channel, respectively.

The second criterion is the convergence of OPE, which ensures that the neglected power corrections in the condensate term remain small and the truncated OPE remains effective. In our calculation, only the four-quark condensate $ \langle \bar{q}q\rangle^2 $ in Eq.~(\ref{f1}) contributes to the expansion of OPE, which means the relative contribution from the condensate $ \langle \bar{q}q\rangle^2 $ needs to be less than $ 30\% $. Additionally, since the Borel parameters $ M_{B_1}^2 $ and $ M_{B_2}^2 $ are not physical parameters, it is necessary to find an optimal window in which the transition form factors exhibit minimal dependence of $ M_{B_1}^2 $ and $ M_{B_2}^2 $.

\begin{figure}[ht]
\centering
\includegraphics[width=7.5cm]{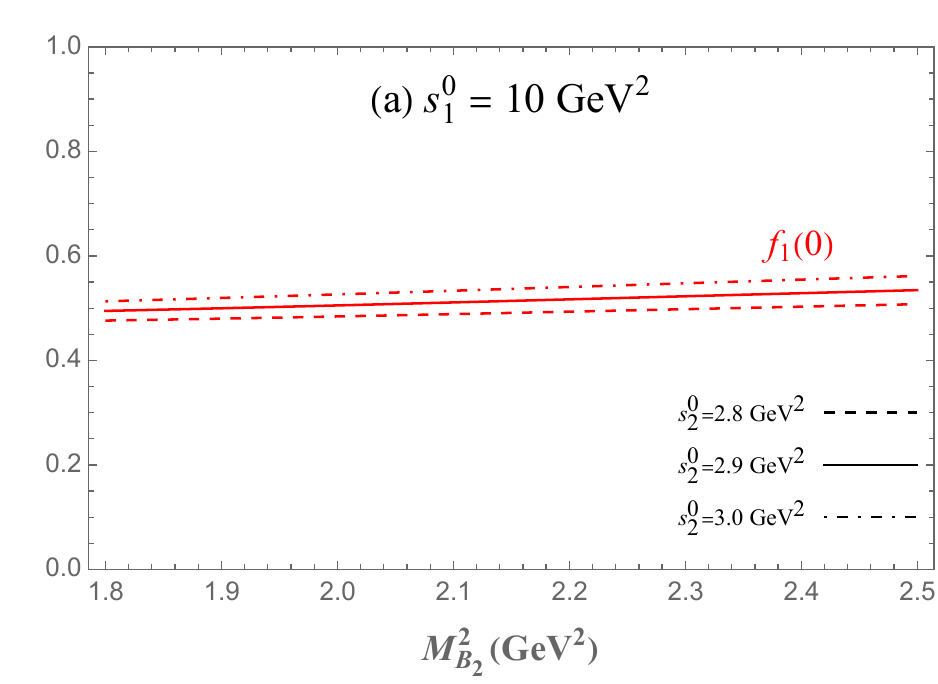}
\includegraphics[width=7.5cm]{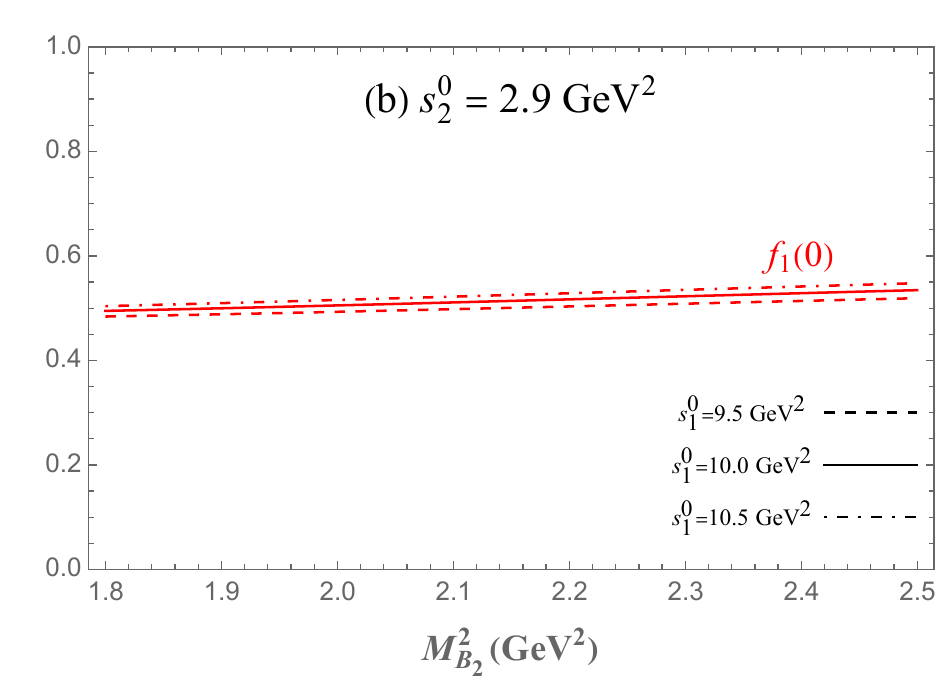}
\includegraphics[width=7.5cm]{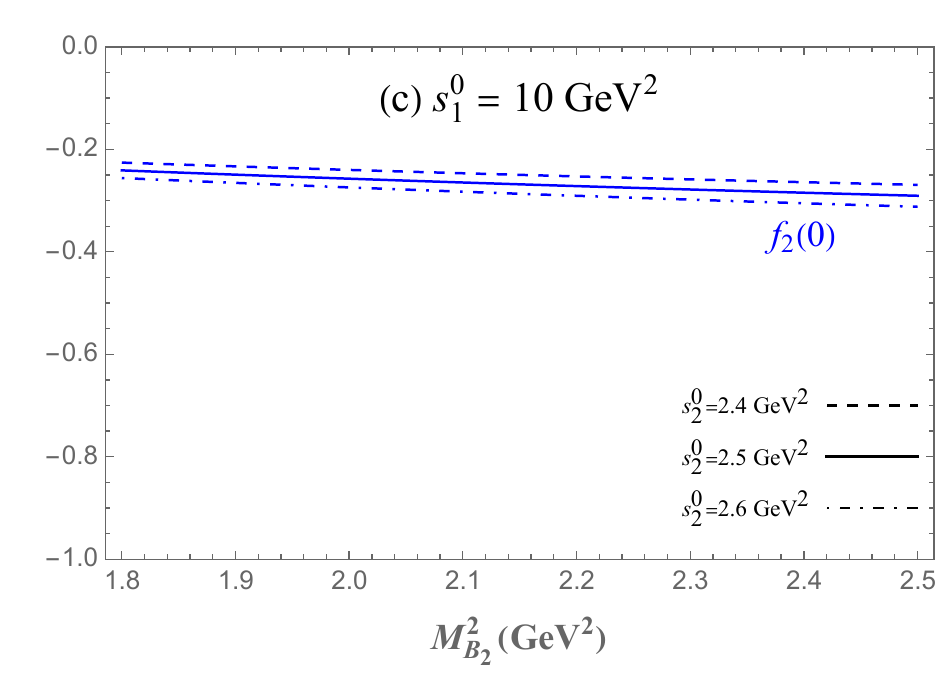}
\includegraphics[width=7.5cm]{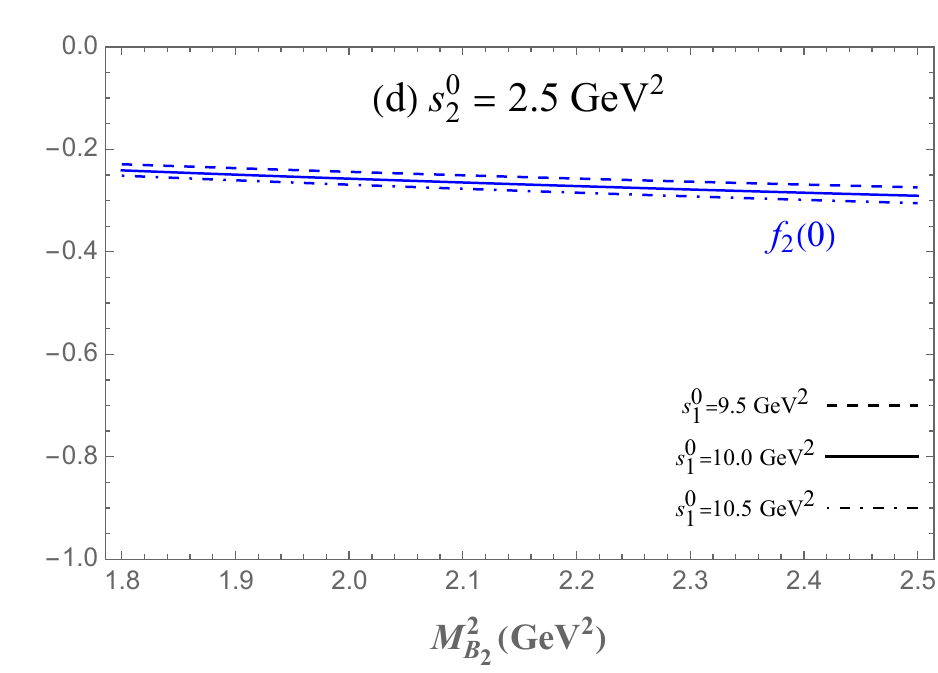}
\caption{The transition form factors $ f_1(0) $ and $ f_2(0) $ as functions of Borel parameter $ M_{B_2}^2 $ for different values of $ s_1^0 $ and $ s_2^0 $.}
\label{fig:f0}
\end{figure}

Through the above preparation, we notice that only one Lorentz structure of $ f_1 $ and $ f_2 $ in Eqs.~(\ref{f1}) and (\ref{f2}) can meet all the above criteria, while we are unable to identify suitable Borel parameters that simultaneously satisfy both criteria for the remaining structures. Then the transition form factors of the semileptonic decay $ \Lambda_c\rightarrow n\ell\nu_\ell $ can be numerically calculated. The dependence of the form factors at the maximum recoil point $ q^2=0 $ with the required range of Borel parameter $ M_{B_2}^2 $ is shown in Fig.~\ref{fig:f0}. In Fig.~\ref{fig:f0}, it can be observed that the variation of $ s_1^0 $ has a negligible effect on $ f_1(0) $ and $ f_2(0) $, whereas the variation of $ s_2^0 $ has a more significant impact. For comparison, we show our results and previous theoretical predictions of transition form factors at maximum recoil point $ q^2=0 $ in Table~\ref{table:f0}. The errors are mainly determined by the uncertainties of the Borel parameters $ M_{B_1}^2 $ and $ M_{B_2}^2 $ and other input parameters listed in Eq.~(\ref{parameter}). In Table.~\ref{table:f0}, our results for $ f_1(0) $, $ f_2(0) $, and $ g_1(0) $ are comparable to other predictions, while there is significant variation for $ g_2(0) $ obtained from different theoretical methods.  In this work, the sign of $ g_2(0) $ aligns with the results from the LCSR~\cite{Khodjamirian:2011jp} and LF approach~\cite{Zhao:2018zcb}, but differs from those derived by other theoretical methods. Further investigations are needed to resolve this discrepancy. Moreover, it is worth mentioning that the results from LCSR~\cite{Khodjamirian:2011jp} are derived using the same interpolating current as in Eq.~(\ref{current-lc}), where the transition form factors at $ q^2=0 $ show a high level of consistency with QCDSR.

\begin{table}[ht]
\centering
\caption{Theoretical predictions for the form factors of the semileptonic decay $ \Lambda_c\rightarrow n \ell \nu_\ell $ at the maximum recoil point $ q^2=0 $ with different approaches.}
\begin{tabular}{lccccccc}
\hline\hline
Method & $ f_1(0) $ &  & $ f_2(0) $ &  & $ g_1(0) $ &  & $ g_2(0) $ \\ \hline
QCDSR   & $0.53 \pm 0.04$       &  & $-0.25\pm 0.03$     &  & $0.53 \pm 0.04$       &  & $-0.25\pm 0.03$       \\
LCSR~\cite{Khodjamirian:2011jp} & $0.59^{+0.15}_{-0.16}$  &  & $-0.43^{+0.13}_{-0.12}$      &  & $0.55^{+0.14}_{-0.15}$       &  & $-0.16^{+0.08}_{-0.05}$      \\
LF~\cite{Zhao:2018zcb}          & 0.513      &  & $-0.266$     &  & 0.443      &  & $-0.034$     \\
CCQM~\cite{Gutsche:2014zna}        & 0.47       &  & $-0.246$     &  & 0.414      &  & 0.073      \\
RQM~\cite{Faustov:2016yza}         & 0.627      &  & $-0.259$     &  & 0.433      &  & 0.118      \\
MBM~\cite{Geng:2020fng}        & 0.40       &  & $-0.22$      &  & 0.43       &  & 0.07       \\
LQCD~\cite{Meinel:2017ggx} & $0.672\pm 0.039$      &  & $-0.321\pm 0.038$     &  & $0.602\pm 0.031$      &  & $0.003 \pm 0.052$      \\
\hline
\end{tabular}
\label{table:f0}
\end{table}

\begin{figure}[ht]
\centering
\includegraphics[width=7.3cm]{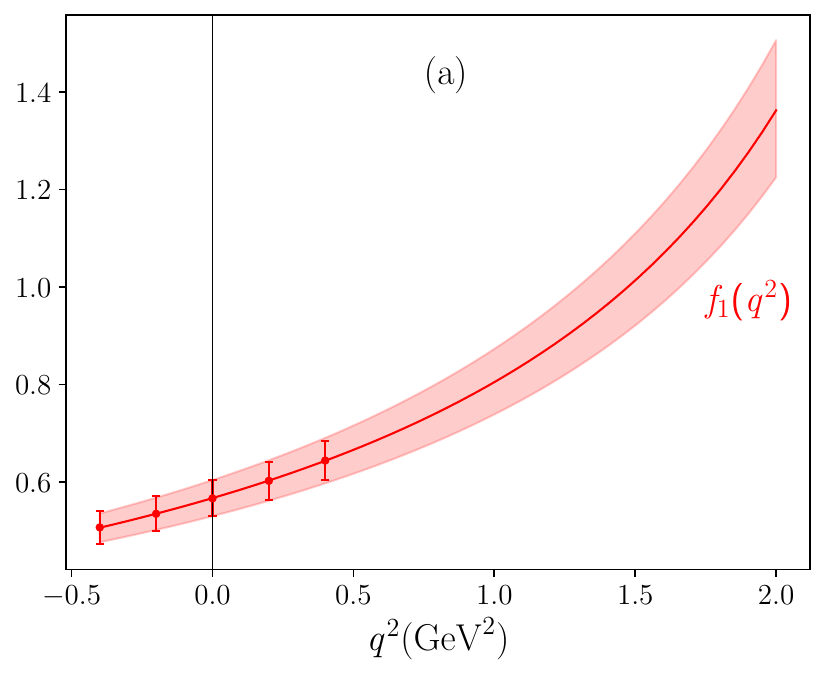}
\includegraphics[width=7.5cm]{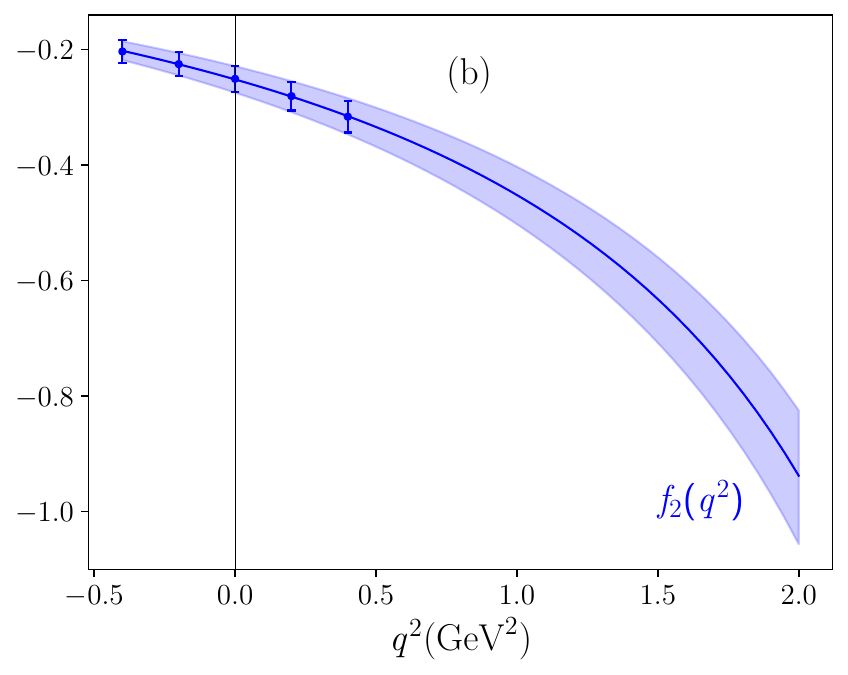}
\caption{The $ q^2 $ dependence of form factors, in which the solid lines represent the central value of fitting parameters listed in Table.~\ref{table:fitting}, while the error bands represent the uncertainties allowed by the fitting parameters. Besides, the symbol point as well as the error bar denote the fitted points and the uncertainties for each of the form factors.}
\label{fig:ft}
\end{figure}

Considering that the QCDSR method is applicable only in the small $ q^2 $ region, and the physical region for $ q^2 $ in the $\Lambda_c\rightarrow n\ell\nu_\ell$ decay extends from $ m_\ell^2 $ to $ (M_{\Lambda_c}-M_n)^2 $, we employ a conformal mapping $ q^2\rightarrow z $ and $ z $-series parametrization to extrapolate the obtained values to the entire physical region. Specifically, we utilize the $ z $-series parametrization in the BCL version proposed in~\cite{Bourrely:2008za}. The mapping transform is expressed as follows:
\begin{align}
\label{z-series}
z(q^2,t_0)=\frac{\sqrt{t_+-q^2}-\sqrt{t_+-t_0}}{\sqrt{t_+-q^2}+\sqrt{t_+-t_0}},
\end{align}
where $ t_{\pm}= (M_{\Lambda_c}\pm M_n)^2$, and $ t_0=t_+-\sqrt{t_+-t_-} \sqrt{t_+-t_{min}}$ is chosen to maximally reduce the interval of $ z $ after mapping $ q^2 $ to $ z $ with the interval $ t_{min} < q^2 < t_- $~\cite{Khodjamirian:2011jp, Huang:2022lfr}. In the numerical analysis, we choose $ t_{min}=-0.4 \,\text{GeV} $. Moreover, the following parametrization is adopted,
\begin{align}
\label{BCL}
f_i(q^2)=\frac{f_i(0)}{1-q^2/(m_{pole})^2}\Bigl\{1+a_1(z(q^2,t_0)-z(0,t_0))\Bigr\}.
\end{align}
Here, $ a_1 $ is a fitting parameter. $ f_i(0) $ represents the value of form factors at $ q^2=0 $, which is also treated as a fitted parameter here. For the pole masses, we employ $ m_{pole}=m_{D^+}=1.87\,\text{GeV} $~\cite{ParticleDataGroup:2022pth} for the $ \Lambda_c\rightarrow n \ell \nu_\ell $ decay mode. To ascertain the central values, uncertainties, and correlation coefficients of the fitted parameters $ f_i(0) $ and $ a_1 $ for each form factor, we begin by generating a set of QCDSR data points for these form factors. The QCDSR data points for each form factor are computed at $ q^2=\{-0.4,-0.2,0,0.2,0.4\} $, utilizing a total of $ N=500 $ ensembles of input parameter sets, which encompass $ M_{B_2}^2 $, $ s_1^0 $, $ s_2^0 $, and other input parameters such as the quark condensate $ \langle \bar{q}q\rangle $. The input parameter values are distributed randomly according to a multivariate joint distribution~\cite{SentitemsuImsong:2014plu}. We subsequently perform a fitting of the $ z $-series expansion to $ f_1(q^2) $ and $ f_2(q^2) $ in order to obtain the fitting parameters, namely $ f_i(0) $ and $ a_1 $, along with the correlation coefficients $\rho$ between them. The fitting results are presented in Table.~\ref{table:fitting}, where the values of $ f_i(0) $ obtained from the fitting procedure are consistent with our directly calculated results given in Table.~\ref{table:f0}. The $ q^2 $ dependence of form factors is shown in Fig.~\ref{fig:ft}.

\begin{table}[ht]
\centering
\caption{Fitted parameters and the correlation coefficient $ \rho $ between them for $ f_1(q^2) $ and $ f_2(q^2) $ using the $ z $-series parametrization in Eq.~(\ref{BCL})}
\begin{tabular}{cclclclc}
\hline\hline
& $ f_i(0) $ &  & $ a_1 $ &  & $ \rho $ \\ \hline
$ f_1 $   & $ 0.57\pm 0.04 $  &  & $ -0.24\pm 0.76 $     &  & $ 0.38 $             \\
$ f_2 $   & $-0.25\pm 0.02 $  &  & $ -10.99 \pm 0.96 $     &  &$ -0.53 $            \\
\hline
\end{tabular}
\label{table:fitting}
\end{table}

\color{black}

After deriving the $ q^2 $ dependence of transition form factors, the branching fractions and the relevant decay asymmetry observables of semileptonic decay $ \Lambda_c\rightarrow n\ell\nu_\ell $ can be analyzed. To facilitate this analysis, it is convenient to introduce the helicity amplitudes, which provide a more intuitive understanding of the physical pictures and simplify the expressions when discussing the asymmetries of the decay processes. The relations between helicity amplitudes and the form factors are as follows~\cite{Li:2021qod, Huang:2022lfr, Shi:2019hbf}:
\begin{align}
& H_{\frac{1}{2}, 0}^V= \frac{\sqrt{Q_{-}}}{\sqrt{q^2}}\big(M_{+} f_1(q^2)-\frac{q^2}{M_{\Lambda_c}} f_2(q^2)\big), \quad H_{\frac{1}{2}, 0}^A= \frac{\sqrt{Q_{+}}}{\sqrt{q^2}}\big(M_{-} g_1(q^2)+\frac{q^2}{M_{\Lambda_c}} g_2(q^2)\big), \nonumber\\
& H_{\frac{1}{2}, 1}^V= \sqrt{2 Q_{-}}\big(-f_1(q^2)+\frac{M_{+}}{M_{\Lambda_c}} f_2(q^2)\big), \quad H_{\frac{1}{2}, 1}^A= \sqrt{2 Q_{+}}\big(-g_1(q^2)-\frac{M_{-}}{M_{\Lambda_c}} g_2(q^2)\big), \nonumber\\
& H_{\frac{1}{2}, t}^V= \frac{\sqrt{Q_{+}}}{\sqrt{q^2}}\big(M_{-} f_1(q^2)+\frac{q^2}{M_{\Lambda_c}} f_3(q^2)\big), \quad H_{\frac{1}{2}, t}^A= \frac{\sqrt{Q_{-}}}{\sqrt{q^2}}\big(M_{+} g_1(q^2)-\frac{q^2}{M_{\Lambda_c}} g_3(q^2)\big).
\end{align}
Here, $ H_{\lambda^\prime, \lambda_W}^{V(A)} $ is the helicity amplitudes for weak transitions induced by vector and axial-vector currents, where $ \lambda^\prime $ and $ \lambda_W $ represent the helicity of the neutron and the $ W $ boson, respectively. $ Q_{\pm} $ is defined as $ Q_{\pm} = M_{\pm}^2-q^2 $ and $ M_{\pm}=M_{\Lambda_c}\pm M_n $. The negative helicity amplitudes can be derived using the following relations:
\begin{align}
H_{-\lambda^\prime,-\lambda_W}^V=H_{\lambda^\prime, \lambda_W}^V, \quad  \quad H_{-\lambda^\prime,-\lambda_W}^A=-H_{\lambda^\prime, \lambda_W}^A.
\end{align}
Then the total helicity amplitudes can be obtained,
\begin{align}
H_{\lambda^\prime, \lambda_W}=H_{\lambda^\prime, \lambda_W}^V-H_{\lambda^\prime, \lambda_W}^A.
\end{align}

With the above helicity amplitudes, the differential distribution of $ \Lambda_c\rightarrow n\ell\nu_\ell $ can be expressed as~\cite{Faustov:2016yza, Shi:2019hbf}
\begin{align}
\label{differential}
\frac{d \Gamma\left(\Lambda_c\rightarrow n\ell\nu_\ell\right)}{d q^2}=\frac{G_F^2\left|V_{c d}\right|^2 q^2\sqrt{Q_+Q_-}}{384 \,\pi^3 \,M_{\Lambda_c}^3}(1-\frac{m_\ell^2}{q^2})^2 H_{\text{tot}},
\end{align}
where $ m_{\ell} $ denotes the lepton mass ($ \ell=e,\mu $) and $ H_{\text{tot}} $ is defined as
\begin{align}
H_{\text{tot}} &= \big(1+\frac{m_\ell^2}{2 q^2}\big)\big(H_{\frac{1}{2},1}^2+H_{-\frac{1}{2},-1}^2+H_{\frac{1}{2},0}^2+H_{-\frac{1}{2},0}^2\big)\nonumber\\[3pt]
&+\frac{3\,m_\ell^2}{2 q^2}\big(H_{\frac{1}{2},t}^2+H_{-\frac{1}{2},t}^2\big).
\end{align}

\begin{figure}[t]
\centering
\includegraphics[width=7.3cm]{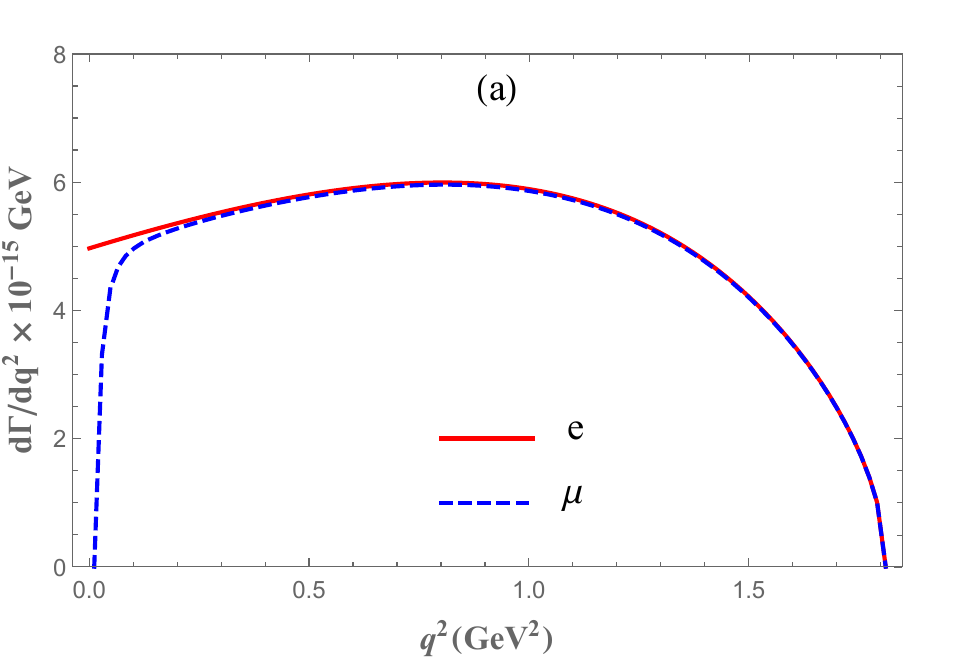}
\includegraphics[width=7.5cm]{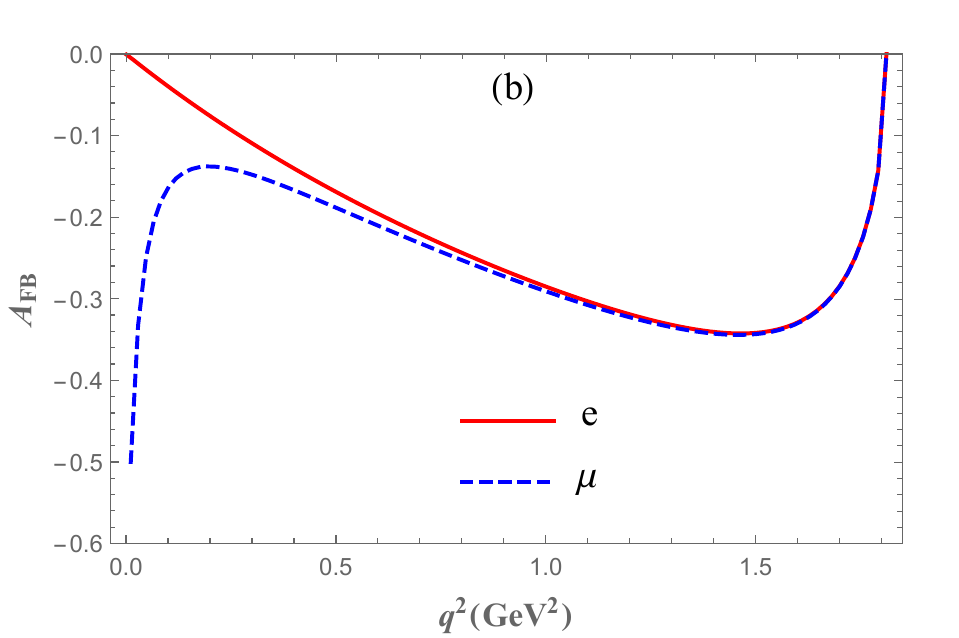}
\includegraphics[width=7.5cm]{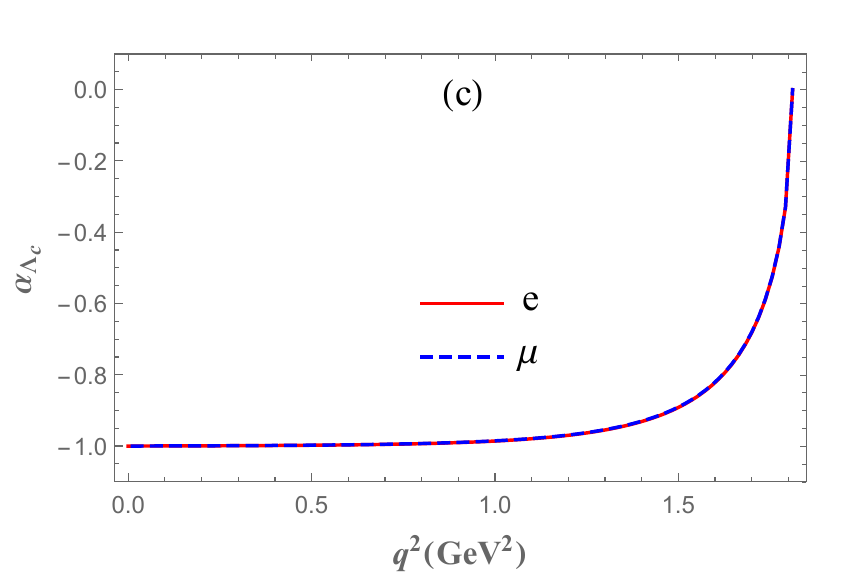}
\caption{The $ q^2 $ dependence of the differential decay width and the relevant decay observables for $ \Lambda_c \rightarrow n \ell \nu_\ell$ semileptonic decay. The red solid line denotes $ \ell=e^+ $, while the blue dashed line denotes $ \ell=\mu^+ $.}
\label{fig:width}
\end{figure}

According to the definition of $ H_{\text{tot}} $, the contribution to the differential decay width from $ f_3(q^2) $ and $ g_3(q^2) $ can be found in the term $ H_{\frac{1}{2},t}^2 $ and $ H_{-\frac{1}{2},t}^2 $, which is clearly suppressed by the factor $ m_\ell^2 $. Hence, we neglect the effect of $ f_3(q^2) $ and $ g_3(q^2) $ in Eq.~(\ref{3ptphe}). In order to obtain the numerical results of differential decay width, the following input parameters related to the decay analysis are taken from the Particle Data Group~\cite{ParticleDataGroup:2022pth}, where
\begin{align}
&G_F=1.166\times 10^{-5}\, \text{GeV}^{-2},\quad |V_{cd}|=0.221\pm0.004,\nonumber\\
&m_e=0.511\,\text{MeV},\quad m_\mu=0.106\,\text{GeV},\quad \tau_{\Lambda_c}=(201.5\pm2.7)\times 10^{-15}\,\text{s}.
\end{align}
Here, the mean lifetime of $ \Lambda_c $, noted as $ \tau_{\Lambda_c} $, is introduced to calculate the branching fractions. We plot the $ q^2 $ dependence of the differential decay width for $ \Lambda_c \rightarrow n \ell \nu_\ell$ semileptonic decay in Fig.~\ref{fig:width}(a) and list the numerical results of branching fractions in Table.~\ref{table:numerical}. It can be found that the branching fractions for $ \Lambda_c \rightarrow n e^+ \nu_e$ semileptonic decay obtained using QCDSR are very close to the results derived by CQM~\cite{Pervin:2005ve}, RQM~\cite{Faustov:2016yza}, $ SU(3) $ flavor symmetry~\cite{Lu:2016ogy}, and MBM~\cite{Geng:2020fng}. As for the $ \Lambda_c \rightarrow n \mu^+ \nu_\mu $ decay mode, we find our results are consistent with RQM~\cite{Faustov:2016yza} and relatively smaller than the Lattice QCD predictions~\cite{Meinel:2017ggx}.

\begin{table}[ht]
\centering
\caption{Theoretical predictions of branching fractions, the forward-backward asymmetry, and the asymmetry parameter for $ \Lambda_c\rightarrow n \ell \nu_\ell $ and $ \Lambda_c\rightarrow \Lambda \ell \nu_\ell $ semileptonic decay with different methods.}
\begin{tabular}{llcccc}
\hline\hline
Channel  & Method  &  & $ \mathcal{B}(\%) $ & $\langle A_{FB} \rangle$ & $\langle \alpha_{\Lambda_{c}} \rangle$  \\ \hline
\multicolumn{1}{l}{\multirow{1}{*}{$\Lambda_c\rightarrow n e^+ \nu_e$}} & QCDSR & & $ 0.281\pm 0.056 $  & $ -0.23\pm 0.01 $ & $ -0.93\pm 0.03 $ \\
\multicolumn{1}{c}{} & LF~\cite{Zhao:2018zcb} & & $ 0.201 $ &  &   \\
\multicolumn{1}{c}{} & CCQM~\cite{Gutsche:2014zna} & & $ 0.207 $ & $-0.236$   &    \\
\multicolumn{1}{c}{} & CQM~\cite{Pervin:2005ve} & & $ 0.270 $ &  &    \\
\multicolumn{1}{c}{} & RQM~\cite{Faustov:2016yza} & & $ 0.268 $ & $-0.251$   & $ -0.91 $  \\
\multicolumn{1}{c}{} & $ SU(3) $~\cite{Lu:2016ogy} & &  $ 0.289\pm 0.035 $ &  &   \\
\multicolumn{1}{c}{} & $ SU(3) $~\cite{Geng:2019bfz} & &  $ 0.51\pm 0.04 $ & & $ -0.89\pm 0.04 $    \\
\multicolumn{1}{c}{} & MBM~\cite{Geng:2020fng} & &  $ 0.279 $ &  & $ -0.87 $    \\
\multicolumn{1}{c}{} & LFCQM~\cite{Geng:2020fng} & &  $ 0.36\pm 0.15 $ & & $ -0.96\pm 0.04 $    \\
\multicolumn{1}{c}{} & LQCD~\cite{Meinel:2017ggx} & & $ 0.410\pm 0.026 $ &  &    \\
\hline
\multicolumn{1}{l}{\multirow{1}{*}{$\Lambda_c\rightarrow n \mu^+ \nu_\mu$}} & QCDSR & & $ 0.275\pm 0.055 $ & $ -0.25\pm 0.02 $ & $ -0.93\pm 0.03 $ \\
\multicolumn{1}{c}{} & CCQM~\cite{Gutsche:2014zna} & & $ 0.202$ & $-0.260$   &    \\
\multicolumn{1}{c}{} & RQM~\cite{Faustov:2016yza} & & $ 0.262$& $-0.276$   & $ -0.90 $    \\
\multicolumn{1}{c}{} & LQCD~\cite{Meinel:2017ggx} & & $ 0.400\pm 0.026 $ &   &    \\
\hline
\multicolumn{1}{l}{\multirow{1}{*}{$\Lambda_c\rightarrow \Lambda e^+ \nu_e$}} & QCDSR & & $ 3.49\pm 0.65 $ & $ -0.20\pm 0.01 $ & $ -0.90\pm 0.03 $ \\
\multicolumn{1}{c}{} & LQCD~\cite{Meinel:2016dqj} & & $ 3.80\pm 0.22  $ & $ -0.20\pm 0.06 $ & $ -0.87\pm 0.10 $  \\
\multicolumn{1}{c}{} & Exp~\cite{BESIII:2022ysa, BESIII:2023jxv} & & $ 3.56\pm 0.11\pm 0.07  $ & $ -0.24\pm 0.03 $ & $ -0.86\pm 0.04 $  \\
\hline
\multicolumn{1}{l}{\multirow{1}{*}{$\Lambda_c\rightarrow \Lambda \mu^+ \nu_\mu$}} & QCDSR & & $ 3.37\pm 0.54 $ & $ -0.24\pm 0.01 $ & $ -0.90\pm 0.02 $ \\
\multicolumn{1}{c}{} & LQCD~\cite{Meinel:2016dqj} & & $ 3.69\pm 0.22  $ & $ -0.17\pm 0.07 $ & $ -0.87\pm 0.10 $  \\
\multicolumn{1}{c}{} &Exp~\cite{BESIII:2022ysa, BESIII:2023jxv}& & $ 3.48\pm 0.17  $ & $ -0.22\pm 0.04 $ & $ -0.94\pm 0.08 $  \\
\hline
\end{tabular}
\label{table:numerical}
\end{table}

In addition, two relevant decay asymmetry observables, e.g., the leptonic forward-backward asymmetry ($ A_{FB} $) and the asymmetry parameter ($ \alpha_{\Lambda_{c}} $) are defined as~\cite{Faustov:2016yza, Geng:2022fsr, BESIII:2023jxv},
\begin{align}
\label{AFB}
A_{F B}(q^2)&=\frac{\frac{d \Gamma}{d q^2}(\text{forward})-\frac{d \Gamma}{d q^2}(\text{backward})}{\frac{d \Gamma}{d q^2}} \nonumber\\
&=\frac{3}{4}\frac{H_{\frac{1}{2}, 1}^2-H_{-\frac{1}{2}, -1}^2-2\frac{m_\ell^2}{q^2}(H_{\frac{1}{2},0}H_{\frac{1}{2},t}+H_{-\frac{1}{2},0}H_{-\frac{1}{2},t})}{H_{\text{tot}}}, \\[5pt]
\label{PB}
\alpha_{\Lambda_{c}}(q^2)&=\frac{d \Gamma^{\lambda^\prime=\frac{1}{2}} / d q^2-d \Gamma^{\lambda^\prime=-\frac{1}{2}} / d q^2}{d \Gamma^{\lambda^\prime=\frac{1}{2}} / d q^2+d \Gamma^{\lambda^\prime=-\frac{1}{2}} / d q^2},
\end{align}
where
\begin{align}
\frac{d \Gamma^{\lambda^\prime=\frac{1}{2}}}{d q^2}= &\, \frac{4 m_l^2}{3 q^2}\big(H_{\frac{1}{2}, 1}^2+H_{\frac{1}{2}, 0}^2+3 H_{\frac{1}{2}, t}^2\big)+\frac{8}{3}\big(H_{\frac{1}{2}, 1}^2+H_{\frac{1}{2}, 0}^2\big),\\[5pt]
\frac{d \Gamma^{\lambda^\prime=-\frac{1}{2}}}{d q^2}= &\, \frac{4 m_l^2}{3 q^2}\big(H_{-\frac{1}{2}, -1}^2+H_{-\frac{1}{2}, 0}^2+3 H_{-\frac{1}{2}, t}^2\big)+\frac{8}{3}\big(H_{-\frac{1}{2}, -1}^2+H_{-\frac{1}{2}, 0}^2\big).
\end{align}

The $ q^2 $ dependence of the decay asymmetry observables are plotted in Figs.~\ref{fig:width}(b) and \ref{fig:width}(c). In Fig.~\ref{fig:width}, we can see the dependence of the differential decay width and the decay asymmetry observables on the lepton mass is consistent near the zero recoil region $ q^2=(M_{\Lambda_c}-M_n)^2 $. However, near the maximum recoil point $ q^2=m_\ell^2 $, the behavior of the differential decay width and the leptonic forward-backward asymmetry is significantly different. The leptonic forward-backward asymmetry is going to $ 0 $ for the $ \Lambda_c\rightarrow n e^+ \nu_e $ decay mode and to $ -0.5 $ for the $ \Lambda_c\rightarrow n \mu^+ \nu_\mu $ decay mode at $ q^2=m_\ell^2 $. This character agrees with Ref.~\cite{Faustov:2016yza}. As for the asymmetry parameter, it varies from $ \alpha_{\Lambda_c}=-1 $ to $ \alpha_{\Lambda_c}=0 $ as the $ q^2 $ increases from zero to $ q^2_{max}$. Besides, it is almost indistinguishable throughout the entire physical region, which is also consistent with the findings in Ref.~\cite{Faustov:2016yza}. We also present the mean values of the relevant decay asymmetry observables in Table~\ref{table:numerical}, which are obtained by separately integrating the numerators and denominators in Eqs.(\ref{AFB}) and (\ref{PB}) over the physical region of $q^2$. From Table.~\ref{table:numerical}, it can be observed that our results for $ \langle A_{FB} \rangle $ and $ \langle \alpha_{\Lambda_{c}} \rangle $ agree with the previous theoretical predictions. Future experiments measuring these observables and comparing them with the predictions of the present study will contribute to our understanding of the relevant decay channels and the internal structures of baryons. In addition, the possibility of new physics effects beyond the standard model can be explored through these observables~\cite{Li:2021qod, Azizi:2019tcn}.

By replacing the second $ d $ quark in the current~(\ref{current-n}) with a strange quark and applying the same analysis procedure, we also investigate the semileptonic decay mode $ \Lambda_c \rightarrow \Lambda \ell \nu_\ell $. The relevant input parameters are $ m_s = 93.4^{+8.6}_{-3.4} \,\text{MeV} $~\cite{ParticleDataGroup:2022pth}, $ \lambda_{\Lambda} = 0.0208 \,\text{GeV}^3 $~\cite{Huang:1998ek}, $ M_{\Lambda}=1.116\,\text{GeV} $~\cite{ParticleDataGroup:2022pth}, $ |V_{cs}|=0.975\pm 0.006 $~\cite{ParticleDataGroup:2022pth}, and $ m_{pole}=m_{D_s^+}=1.97\,\text{GeV} $~\cite{ParticleDataGroup:2022pth}. As this decay mode has been thoroughly investigated both theoretically and experimentally, we solely provide the numerical results of branching fractions and decay asymmetry observables in Table.~\ref{table:numerical} as a validation of the QCDSR method. It is evident that the branching fractions, the forward-backward asymmetry, and the asymmetry parameter obtained through QCDSR for the semileptonic decay $ \Lambda_c \rightarrow \Lambda \ell \nu_\ell $ are in excellent agreement with Lattice QCD results~\cite{Meinel:2016dqj} and experimental data~\cite{BESIII:2022ysa, BESIII:2023jxv}.

\section{Conclusions}
In this work, we calculate the weak transition form factors of $ \Lambda_c \rightarrow n \ell \nu_\ell $ semileptonic decay in the framework of QCD sum rules. The analytic results of the transition form factors are obtained through the analysis of the three-point correlation functions and the application of Cutkosky cutting rules. The numerical results for the form factors at the maximum recoil region point $ q^2=0 $ are computed and compared with other methods. In order to extend the form factors to the full physical region, we utilize a $ z $-series parametrization that adequately captures the $ q^2 $ dependence of the form factors, ensuring a smooth extrapolation.

Based on the obtained form factors, we predict the branching fractions to be $ \mathcal{B}(\Lambda_c\rightarrow n e^+ \nu_e)= (0.281\pm 0.056)\% $ and $ \mathcal{B}(\Lambda_c\rightarrow n \mu^+ \nu_\mu)= (0.275\pm 0.055)\% $, which will provide important information to determine the value of the CKM matrix element $ |V_{cd}| $. Moreover, the mean values of the leptonic forward-backward asymmetry $ \langle A_{FB} \rangle $ and the asymmetry parameter $ \langle \alpha_{\Lambda_{c}} \rangle $ are also given, which will play a crucial role in probing potential new physics effects beyond the standard model. Although there are still no experimental data for $ \Lambda_c\rightarrow n\ell \nu_\ell $ semileptonic decay to date, considering the recent experimental progress of $ \Lambda_c $ decay modes involving the neutron final state, we believe our predicted results can be tested by the future experiments at BESIII, BELLEII, and LHCb.

Finally, we analyze the semileptonic decay mode $ \Lambda_c \rightarrow \Lambda \ell \nu_\ell $. Our results exhibit a strong agreement with the experimental data, indicating that the QCDSR calculation can be applied to other charmed baryons, such as $ \Xi_c^{+(0)} $. Furthermore, there is still potential for further improvement in this method. The relatively large errors in the branching fractions compared with the experimental data suggest the necessity for further refinement. One possible approach to address this issue is to calculate the contributions from radiation corrections, although this presents a significant challenge in the application of QCD sum rules.

\vspace{0.5cm}
{\bf Acknowledgments}

We thank K.S. Huang, L. Tang, B.D. Wan, and D.S. Ye for their meaningful discussions. This work was supported in part by National Natural Science Foundation of China(NSFC) under Grants No. 11975236 and No. 12235008, and the University of Chinese Academy of Sciences.


\end{document}